\documentclass[sigconf,natbib]{acmart}
\usepackage{times}
\usepackage{graphicx}
\usepackage{latexsym}
\usepackage{subcaption}
\usepackage{multirow}
\usepackage{makecell}
\usepackage{balance}
\usepackage{booktabs}
\usepackage{array}
\usepackage{enumitem}
\usepackage{amsmath}
\usepackage{stfloats}

\usepackage{amssymb}
\usepackage[show]{chato-notes}

\AtBeginDocument{%
  \providecommand\BibTeX{{%
    \normalfont B\kern-0.5em{\scshape i\kern-0.25em b}\kern-0.8em\TeX}}}

 \providecommand\BibTeX{{%
  Bib\TeX}}

\copyrightyear{2024}
\acmYear{2024}
\setcopyright{rightsretained}
\begin{document}

\title{Unsupervised dense retrieval with conterfactual contrastive learning}


\author{Haitian Chen}
\orcid{0009-0006-7208-9675}
\affiliation{%
  \institution{Tsinghua University}
  \city{Beijing}
  \country{China}
  \postcode{100084}
}
\email{chenhaitian233@gmail.com}

\author{Qingyao Ai}
\orcid{0000-0002-5030-709X}
\affiliation{%
  \institution{Tsinghua University}
  \city{Beijing}
  \country{China}
  \postcode{100084}
}
\email{aiqy@tsinghua.edu.cn}

\author{Xiao Wang}
\affiliation{%
  \institution{University of Glasgow}
  \city{Glasgow}
  \country{United Kingdom}
}
\email{x.wang.8@research.gla.ac.uk}


\author{Yiqun Liu}
\orcid{0000-0002-0140-4512}
\authornote{Corresponding author}
\affiliation{%
  \institution{Tsinghua University}
  \city{Beijing}
  \country{China}
  \postcode{100084}
}
\email{yiqunliu@tsinghua.edu.cn}

\author{Fen Lin}
\affiliation{%
  \institution{WeChat Search Application Department, Tencent}
  \country{China}
}
\email{felicialin@tencent.com}

\author{QIN LIU}
\affiliation{%
  \institution{WeChat Search Application Department, Tencent}
  \country{China}
}
\email{stenliu@tencent.com}


\renewcommand{\shortauthors}{Haitian Chen et al.}

\begin{abstract}

Efficiently retrieving a concise set of candidates from a large document corpus remains a pivotal challenge in Information Retrieval (IR). 
Neural retrieval models, particularly dense retrieval models built with transformers and pretrained language models, have been popular due to their superior performance. 
However, criticisms have also been raised on their lack of explainability and vulnerability to adversarial attacks.
In response to these challenges, we propose to improve the robustness of dense retrieval models by enhancing their sensitivity of fine-graned relevance signals. 
A model achieving sensitivity in this context should exhibit high variances when documents' key passages determining their relevance to queries have been modified, while maintaining low variances for other changes in irrelevant passages.
This sensitivity allows a dense retrieval model to produce robust results with respect to attacks that try to promote documents without actually increasing their relevance.
It also makes it possible to analyze which part of a document is actually relevant to a query, and thus improve the explainability of the retrieval model.
Motivated by causality and counterfactual analysis, we propose a series of counterfactual regularization methods based on game theory and unsupervised learning with counterfactual passages.
Experiments show that, our method can extract key passages without reliance on the passage-level relevance annotations. 
Moreover, the regularized dense retrieval models exhibit heightened robustness against adversarial attacks, surpassing the state-of-the-art anti-attack methods.

\end{abstract}




\maketitle

\section{Introduction}
How to efficiently retrieve a concise set of candidates from an extensive pool of documents is a fundamental challenge in Information Retrieval (IR).
In recent years, neural retrieval models have attracted considerable attention for their superior performance. 
Dense retrieval models, in particular, have not only achieved state-of-the-art retrieval results but have also demonstrated comparable efficiency to traditional systems based on term matching models and inverted indexes~\cite{robertson2009probabilistic, ramos2003using}.

However, the susceptibility of modern retrieval models to adversarial attacks poses a critical limitation. 
Deep learning-based models, including dense retrieval models, are vulnerable to the injection of input "noise," leading to significant variations in the treatment of documents with imperceptible differences to humans within the retrieval system. 
Previous studies~\cite{DBLP:journals/corr/abs-2306-12756, DBLP:conf/cikm/Liu0GR0FC23} have illustrated how replacing words with specific synonyms can deceive dense retrieval models, manipulating the ranking position of a target document regardless of its actual relevance to the query. 
Guided by these findings, we propose a training task to aid dense retrieval models in learning the distinction between positive documents and adversarial documents, thereby enhancing the models' robustness.

This paper posits that a key aspect of mitigating vulnerability is to augment the sensitivity of dense retrieval models to fine-grained relevance matching signals. 
The sensitivity of a particular input data segment generally reflects its significance within the neural network model.
Ideally, a retrieval model should adeptly capture both the overall relevance of a document and the role each part of the document plays in fulfilling the information need of a given query.
Therefore, we argue that a retrieval model demonstrates sensitivity to fine-grained relevance matching if its ranking scores for a specific query-document pair exhibit:
(1) high variances when the key passages of the document that satisfies the query need are modified, and
(2) low variances when the irrelevant passages of the document with respect to the current query are changed.
The former indicates that the model should be sensitive to passages in a document that potentially determine its relevance to the query, while the later indicates that the model should be robust and stable with respect to changes on passages that are unimportant to the query.
This approach makes the model more resistant to adversarial attacks, as any attacks on terms or passages that don't determine the relevance of a document would not lead to significant changes in model outputs.

To address this limitation, we propose a counterfactual passage extraction method based on the Shapley value~\cite{winter2002shapley}.
The objective is to identify the key passage of each document, termed the counterfactual passage, determining its relevance to a given query. 
Subsequently, we introduce several unsupervised learning tasks based on these counterfactual passages to enhance the learning process of dense retrieval models. 
Experimental results demonstrate that our method and model can learn to extract key passages influencing the relevance between a document and a query without requiring passage-level relevance annotations. 
Additionally, the regularized dense retrieval models exhibit robustness against adversarial attacks, surpassing even state-of-the-art anti-attack methods designed specifically based on attack properties.

\section{Related Work}
\subsection{Dense Retieval}
Based on whether term-level interactions are modeled between query and documents beyond their final encodings, Neural IR (Neu-IR) methods can be categorized into representation-based or interaction-based~\cite{DBLP:conf/cikm/GuoFAC16, DBLP:journals/ftir/MitraC18}. 

Interaction-based models enjoy fine-grained modeling of term-level interactions between query and documents; thus they are typically more effective though more expensive and usually used as re-rankers since that requires scoring candidate documents according to the given query~\cite{DBLP:conf/cikm/GuoFAC16, DBLP:conf/wsdm/DaiXC018, DBLP:conf/emnlp/HuiYBM17, DBLP:conf/sigir/MacAvaneyYCG19, DBLP:journals/corr/abs-1901-04085, DBLP:conf/sigir/XiongDCLP17}. Representation-based ones, often encode queries and documents as low-dimensional dense representations without explicit term-level matches. The representation-based models can achieve more efficient retrieval with the document representation precomputing and the support of approximate nearest neighbor (ANN)~\cite{DBLP:journals/corr/abs-2004-13969, DBLP:conf/emnlp/KarpukhinOMLWEC20, DBLP:conf/sigir/KhattabZ20, DBLP:conf/acl/LeeCT19, DBLP:journals/tacl/LuanETC21, DBLP:conf/iclr/XiongXLTLBAO21}. The representation-based models help to achieve an efficient dense retrieval, benefiting many downstream tasks by providing more accurate evidence, such as fact verification, conversational dense retrieval, and open domain question answering~\cite{DBLP:conf/emnlp/KarpukhinOMLWEC20, DBLP:conf/nips/LewisPPPKGKLYR020, DBLP:conf/sigir/Qu0CQCI20, DBLP:conf/iclr/XiongLIDLWMY0KO21}.

\subsection{Counterfactual Learning}
The counterfactual inference has been applied to representation learning to obtain fair representations~\cite{DBLP:conf/nips/KusnerLRS17} in various domains such as image classification~\cite{DBLP:conf/icml/GoyalWEBPL19} and vision-language tasks~\cite{DBLP:conf/emnlp/LiangJHZ20, DBLP:conf/cvpr/NiuTZL0W21}. The key idea behind counterfactual learning is to train a model that is invariant to specific aspects of the input data~\cite{DBLP:conf/icml/JohanssonSS16, DBLP:conf/nips/KusnerLRS17}.

Existing work on counterfactual learning for natural language expands upon this idea and aims to learn robust representations of text data by capturing causal features while mitigating spurious correlations~\cite{DBLP:conf/emnlp/ChoiPYH20, DBLP:conf/emnlp/LiangJHZ20, DBLP:conf/aaai/Tian0ZX22}. A general approach is to apply contrastive learning to differentiate factual (or positive) samples from the counterfactual samples, which are minimally dissimilar and of different labels. Such an approach usually comes with dedicated masking strategies to minimize causal associations in counterfactual samples, and applying counterfactual learning with synthetic data has been shown to yield robust representations. However, few studies have delved into the effect of counterfactual learning on retrieval tasks.

\subsection{Robustness of Dense Retrievers}
In recent years, model robustness has attracted attention in various fields~\cite{DBLP:journals/tmm/WangGLZWN22, DBLP:journals/tkde/LiangLZTWYDL24, DBLP:journals/tmm/WangGLWCN23, DBLP:conf/sigir/0006MLLTWZL23}, including IR~\cite{DBLP:journals/corr/abs-2306-12756, DBLP:conf/cikm/Liu0GR0FC23}. 
In the context of robustness, adversarial attacks aim to discover human-imperceptible perturbations that can deceive neural networks~\cite{DBLP:journals/corr/SzegedyZSBEGF13}. 
Wu et al.~\cite{DBLP:journals/tois/WuZGRFC23} introduced the WSRA method of attacking black-box NRMs using word substitution. 
This study revealed the serious vulnerability of NRMs to synonym substitution perturbations. 
As a result, subsequent explorations of attack against NRMs have emerged~\cite{liu2023topic, chen2023towards}, inspired by this pioneering work. 

In response to adversarial attacks, research has proposed various defense strategies to enhance adversarial robustness. These can be generally classified into certified defenses and empirical defenses. 
Certified defenses aim for theoretical robustness against specific adversarial perturbations~\cite{raghunathan2018certified}. 
For instance,~\cite{wu2022certified} introduced a certified defense method that ensures the top-K robustness of NRMs via randomized smoothing. 

Empirical defenses aim to enhance the empirical robustness of models against known adversarial attacks, and this approach has been extensively explored in image classification~\cite{madry2017towards, wang2021convergence} and text classification~\cite{ye2020safer, jia2019certified}. Among these methods, adversarial training emerges as one of the most effective defenses. Adversarial training on adversarial examples remains empirically robust~\cite{cui2021learnable}.

\section{Methodology}
This section describes our methods on counterfactual passage extraction and counterfactual contrastive learning.
We first introduce the preliminaries of the retrieval task (Sec.~\ref{preliminary}) and how to extract counterfactual passages from a document based on a specific retrieval model (Sec.~\ref{counter_method}). Finally, we describe our contrastive learning task based on counterfactual analysis (Sec.~\ref{contrastive}).

\subsection{Preliminary}\label{preliminary}
Given a query $q$ and a document collection $D = \{d_1, \dots, d_n\}$, dense retrievers calculate the relevance score $f(q, d)$ based on the dense representations of the query and document.
In particular, the representations of a query and a document is denoted as $e_q$ and $e_d$, respectively.
Then the similarity score, i.e. $f(q,d)$, of query $q$ and document $d$ can be calculated with their dense representations:
\begin{equation}
    f(q, d) = sim(e_q, e_d),
\end{equation}
where $sim(\cdot)$ is the similarity function, which is used to estimate the relevance between two embeddings.
The dot product and cosine similarity is usually used as the similarity function.

To obtain the relevance score $f(q, d)$, the dense retrievers are typically trained using the triplet training samples, consisted of the query $q$ and its positive (relevant) document $d^+$ and negative (irrelevant) document $d^-$. 
The dense retrieval models are optimised by minimising the following loss function:
\begin{equation}
    L = \sum\limits_q  l(q, d^+, D^-),
\end{equation}
where $D^-$ is the collection of the negative samples (i.e., $d^-$) for the query $q$, 
and $l(q, d^+, D^-)$ is the contrastive training loss function, defined as following:
\begin{equation}
    \label{dpr_loss}
    l(q, d^+, D^-) = - log \frac{e^{f(q, d^+)}}{e^{f(q, d^+)}+\sum\limits_{d^- \in D^-} e^{f(q, d^-)}}.
\end{equation}


\subsection{Counterfactual Passage Extraction}\label{counter_method}
Ideally, a perfect retrieval model should be able to not only estimate the relevance between documents and queries, but also capture the key passages of a document that determine its relevance to each query. 
Therefore, if we remove or alter such key passages, the predicted relevance scores between the query-document pair should be decreased significantly.
Based on this hypothesis, we propose to extract the key passages that determine a document's relevance through counterfactual analysis.
Specifically, we propose a model-agnostic counterfactual passage extraction method based on the concept of Shapley value in Game theory~\cite{winter2002shapley}. 


For a given query $q$ and document $d$, where the document $d$ can be lengthy and thus consists of different passages $P_d = {p_1, \dots, p_n}$.
Each passage may convey varying amounts of information, and thus possess varying importance when measuring the relevance score of a query.
In particular, we denote the relevant document of a query $q$ as $d^+$, and the most important passage that determines $d^+$'s relevance as $p^+$.
We define a counterfactual document $d_i^*$ as the positive document $d^+$ with the $i$-th passage, i.e. $p_i$, changed.
There are multiple types of modifications one could do on $p_i$, and for simplicity, we only implement and test three of them, namely 
\begin{itemize}[leftmargin=*]
    \item \textbf{Deletion}: Removing the passage from the document;
    \item \textbf{Modification}: Randomly altering words or phrases within the passage;
    \item \textbf{Replacement}: Substituting the passage with another random passage from the document.
\end{itemize}
As shown in Section~\ref{sec:result}, deletion yields the best performance among these three, so we use it as the default modification method in the rest of this paper.

Based on the above definition, a naive counterfactual method to find the key passages based on a document retrieval model is to modify each passage separately and examine which can lead to the largest output change (i.e., the fluctuation of the predicted relevance score).
However, as shown in our experiments (discussed in Section~\ref{sec:result}), such methods perform suboptimally on dense retrieval models.
One of the reasons is that, in dense retrieval models, each document and query are encoded separately, and the relevance score of a document is not a simple aggregation of each passage's relevance.
Motivated by the Shapley value in coalitional game theory~\cite{winter2002shapley}, we present a novel counterfactual method to extract key passages for dense retrieval models. 
Specifically, the extraction of key passages can be conceptualized as a coalitional game, where passages collaborate together to enhance the query-document relevance. 
 We assume that the importance of each passage with respect to the relevance between the query-document pair can be reflected by their contribution to the game, which could be measured with the Shapley value. 
Therefore, by ranking passages based on their Shapley value, we can identify the most crucial passage for each query-document pair.
Formally, assuming that we modify passage with \textbf{Deletion}, based on the definition of the Shapley value, we calculate the contribution of a passage using the following formula:
\begin{equation}
    \phi_v(p_i) = \sum\limits_{P \subset P_d \setminus \{p_i\}} \frac{|P|!(n-|P|-1)!}{n!}(v(P\cup \{p_i\}) - v(P)),
\end{equation}
Here, $v$ represents the relevance score function(e.g. cosine similarity), $p_i$ is the target passage to modify, $P_d \setminus \{p_i\}$ denotes the set of passage in $d$ excluding $p_i$, and $\phi_v(p_i)$ signifies the contribution (also known as the Shapley value) of passage $p_i$.
Intuitively, Shapley value computes the change of the document's relevance score by sampling different passage combination from the document after deleting $p_i$. 
If a passage is considered important in a document by a dense retrieval model, its contribution to relevance should be irreplaceable no matter how we combine other passages in the document.

Our proposed method has two advantages.
First, drawing from the coalitional game theory, our method establishes a more robust theoretical foundation for key passage extractions from documents, which improves the interpretability of dense retrieval models. 
Second, more importantly, our Shapley-based method is model-agnostic and adaptable to various dense retrieval models, which serves as the foundation of the proposed counterfactual contrastic learning method described in the following section.

\begin{figure}[tbp]
    \centering
    \includegraphics[width = \linewidth]{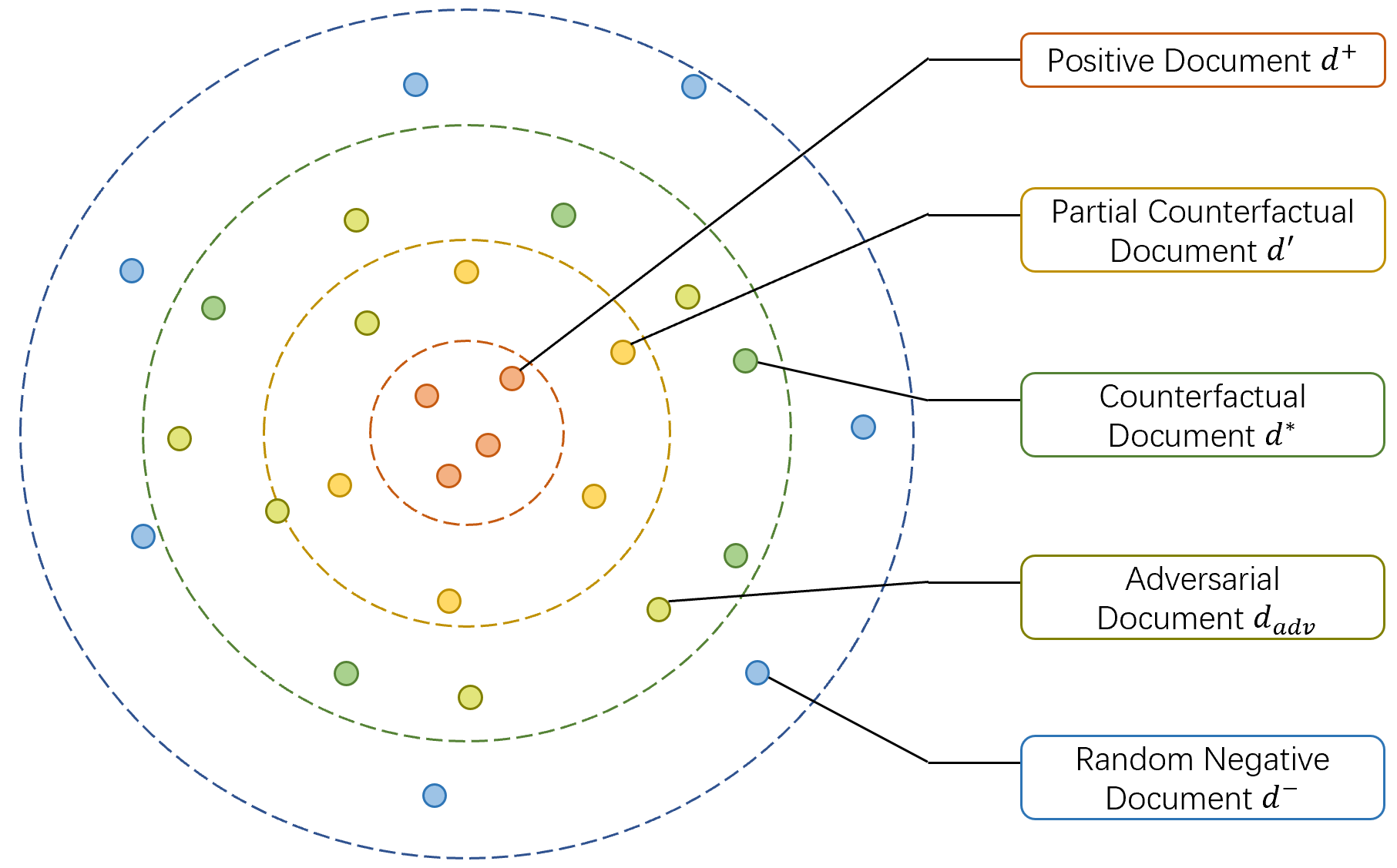}
    \caption{Illustration of the counterfactual and adversarial documents as pivots in the embedding space.}
    \label{fig:3_contrastive_learning_figure}
    \vspace{-0.5cm}
\end{figure}

\subsection{Counterfactual Contrastive Learning }\label{contrastive}

As discussed previously, a robust retrieval model should be sensitive to the changes of key relevance information in a document which being insensitive to modifications on the rest of the document.
If so, the model will be interpreted easily through counterfactual analysis, and immune to adversarial attacks that try to promote certain documents without actually improving their relevance to the query.   
In this section, we propose a counterfactual contrastive learning method based on the counterfactual passage extraction to improve the robustness and relevance sensitivity of dense retrieval models.
Having high relevance sensitivity means that a dense retrieval model could easily distinguish not only positive documents from negative ones, but also counterfactual documents, which modify the key passages of postive documents, from other documents.
To achieve this goal, based on the key passages extracted by Shapley value, we design a pretraining task that requires the dense retrieval model to position different counterfactual documents at proper locations in their embedding space from dense retrievers.
Specifically, we create three types of counterfactual documents with diverse difficulties, which we refer to as the \textit{partial counterfactual}, \textit{full counterfactual}, and \textit{adversarial counterfactual} documents.
The assumptions on the relative preferences between positive documents, negative documents, and counterfactual documents in terms of relevance is depicted in Figure~\ref{fig:3_contrastive_learning_figure}. 

\paragraph{\textbf{Partial/Full Counterfactual Document}}

Formally, for a query-document pair $(q, d^+)$, we define its key passage as $p^+$, and a set of N negative documents  as $D^-$. 
The idea of contrastive learning to train the dense retrieval models to create query and document representations such that the following inequality between positive document $d^+$ and negative document $d^- \in D^-$ holds:
\begin{equation}
    f(q, d^+) > f(q, d^-).
\end{equation}

We define partial counterfactual documents as those with minor modifications from the original positive document $d^+$.
Let a partial counterfactual document be $d'$, then we construct $d'$ by randomly removing one sentence from $p^+$ in $d^+$.
In contrast, we define full counterfactual documents as those with significant differences with $d^+$.
Let a full counterfactual document be $d*$, then we construct $d*$ by removing $p^+$ from $d^+$ directly.
Consequently, we can get the relationship between embedding similarities:
\begin{equation}
    f(q, d^+) > f(q, d') > f(q, d^*).
\end{equation}
Also, since a counterfactual document $d^*$ is a minimally different to the document that retains most of the semantics in $d^+$, $d^*$ can be seen as a pseudo-positive example in the document retrieval. Semantic relevance between $q$ and $d^*$ distinguishes $d^*$ from other negatives $d^-$, which provide noisy contexts with respect to $q$. Thus, the following holds for almost all $d^*$: 
\begin{equation}
    f(q, d^*) > f(q, d^-).
\end{equation}

The idea of partial and full counterfactual documents is to create cases with varying difficulties for the training of dense retrieval. 
Different from traditional hard negative mining techniques, the introduction of our counterfactual documents focus on the modifications of the positive documents, thus are more effective in improving the relevance sensitivity of dense retrieval models instead of their overall retrieval performance.


\paragraph{\textbf{Adversarial construction of Counterfactual Document}}
Other than simply removing sentences and passages from documents, more typical and important document modification cases in practice are those constructed with ill intents to trick search systems, e.g., adversarial attacks. 
Thus, we also introduce a type of counterfactual documents based on adversarial attacks

While the aforementioned partial and full counterfactual documents changes a continuous range text in a local passage, the adversarial counterfactual documents modify different parts of the document globally.
Specifically, we construct the adversarial counterfactual documents $d_{adv}$ as:
\begin{equation}
d_{adv} = arg \max\limits_{d' \in \mathbb{B}(d, \epsilon)} \left( f(q, d') - f(q, d) \right),
\end{equation}
where $\mathbb{B}(d, \epsilon)$ signifies that a ratio of $\epsilon$ words in $d$ are randomly replaced. 
Similarly, based on the difficulties of each counterfactual examples, the following holds for all $d_{adv}$: 
\begin{equation}
    f(q, d^+) > f(q, d_{adv}) > f(q, d^-).
\end{equation}

\paragraph{\textbf{Counterfactual Contrastive Loss}}
Building upon the aforementioned relationships, we propose a counterfactual contrastive learning method for a more robust dense retriever. 
Specifically, we augment the loss function of the dense retriever (Equation \ref{dpr_loss}) by introducing additional counterfactual contrastive loss terms. 
These terms leverage counterfactual documents as pivotal elements between positives and negatives. 
Specifically, these include:
\begin{itemize}

\item \textit{Classic Dense Retrieval Loss:} $L_{cla}$, is a pairwise loss as the classic ranking loss, defined in Equation~\ref{dpr_loss}. 

\item \textit{Counterfactuals as Hard Negatives Loss:} $L_{neg}$, is optimized to maximize the similarity between $(q, d), (q, d')$ while minimizing the similarity between $(q, d'), (q, d^*)$. It imposes the key constraint on $q, d', d^*$ from Equation 6 to discriminate positive documents from counterfactual documents:
\begin{equation}
    L_{neg} = -log \frac{e^{f(q, d^+)}}{e^{f(q, d^+)} + e^{f(q, d')}} -log \frac{e^{f(q, d')}}{e^{f(q, d')} + e^{f(q, d^*)}}.
\end{equation}

\item \textit{Counterfactuals as Pseudo Positives Loss:} $L_{pos}$, is optimized to maximize the relative similarity between $q$ and $d^*$ with respect to negative documents $d_j^-$. The key difference between $L_2$ and $L_3$ is that the counterfactual passage is used as a positive in $L_2$ to retain semantic relevance in the learned embeddings. 
\begin{equation}
    L_{pos} = -log \frac{e(f(q, d^*)}{e^{f(q,d^*)} + \sum_{j=1}^Ne^{f(q, d_j^-)}}.
\end{equation}

\item  \textit{Counterfactuals with Adversarial Attack Loss:}
$L_{adv}$ is optimized to maximize the similarity between $(q, d^+), (q, d_{adv})$ while minimizing the similarity between $(q, d_{adv}), (q, d^*)$.
\begin{equation}
    L_{adv} = -log \frac{e^{f(q, d^+)}}{e^{f(q, d^+)} + e^{f(q, d_{adv})}} -log \frac{e^{f(q, d_{adv})}}{e^{f(q, d_{adv})} + e^{f(q, d^*)}}.
\end{equation}
\end{itemize}

The final loss function L is a weighted sum of all three loss fuctions: 
\begin{equation}
    L = L_{cla} + \alpha (L_{neg} + L_{adv}) + \beta L_{pos},
\end{equation}
where $\alpha, \beta$ are hyperparameters that determine the importance of the terms.
Various strategies exist for setting these hyperparameters based on the documents and passages. In our approach, we explore three different methods:
\begin{itemize}[leftmargin=*]
    \item Relevance score (rel): Setting $\alpha = 1-r, \beta = r, r = f(q, d^*)$.
    \item Shapley value (shapley): Setting $\alpha = 1-s', \beta = s'$, where $s'$ is the normalized shapley value.
    \item Couple learning (CP): Employing the couple learning method~\cite{liu2023perturbation} to dynamically adjust the hyperparameters.
\end{itemize}

\section{Experiments}
\subsection{Dataset Selection and Preprocessing}\label{ssec:dataset}
In all our experiments, we utilize the MSMARCO-doc and MSMARCO-passage datasets~\cite{nguyen2016ms} to evaluate model performance. These datasets comprise extensive anonymized questions sampled from Bing's search query logs and texts extracted from 3,563,535 web pages retrieved by Bing. 
Relevant documents for each query are identified based on the presence of at least one related passage. 
By aligning the MSMARCO-doc dataset with the MSMARCO-passage dataset, we curate matching triples consisting of <query, document, relevant passage>, resulting in a total of 164,190 instances.


\paragraph{\textbf{Key Passage Extraction}}
Since the documents in MSMARCO-doc do not contain passage information, we segment the documents in the dataset using a fixed-size sliding window approach.
We use two window sizes, 64 and 128, with a 50\% overlap.
From the segmented document sections, we identify the segment with over 90\% overlap with the relevant passage in each triple as the positive segments. 
This segment is then annotated as the positive example for both training and evaluation purposes.


Given the requirement for non-overlapping passages in calculating the Shapley value, we employ two methods to obtain the final Shapley value:
\begin{itemize}[leftmargin=*]
    \item \textbf{Non-overlap}: In this approach, only half of the segmented passages, which do not overlap with each other, are taken into account.
    \item \textbf{Merge}: In this approach, the Shapley value of two groups of non-overlapping passages is independently calculated, and the final Shapley value is obtained as the moving average of three consecutive passages.
\end{itemize}

\paragraph{\textbf{Counterfactual Contrastive Learning}}
Given that passage-level relevance labels are often costly and time-consuming to obtain, we explore two training setups to get the relevant passage within the document.
In the first setup, we use the aforementioned shapley-value-based method to extract the key passage within the positive document as the relevant passage. 
In the second setup, we employ a large language model (LLM) to assist with the labeling of relevant passages.
Specifically, we use GPT-3.5 to label the relevant passage in each <query, document> pair, simulating a scenario with limited labeled data.

\subsection{Baselines and Evaluation Metrics}\label{ssec:baselines}
\paragraph{\textbf{Key Passage Extraction}}
We consider two intuitive counterfactual methods for the comparison of model-agnostic key passage extraction methods:
\begin{itemize}[leftmargin=*]
    \item Document Ranking Change ($\delta_{\text{rank}}$): We measure the change in ranking of the modified document compared to the original document in the retrieval model's results. A higher $\delta_{\text{rank}}$ signifies a more crucial passage for the query.
    \item Relevance Score Change ($\delta_{\text{rel}}$): We calculate the differences in the relevance scores assigned by the retrieval model to the original and counterfactual documents. A larger $\delta_{\text{rel}}$ indicates greater passage importance.
\end{itemize}
Based on $\delta_{\text{rank}}$ and $\delta_{\text{rel}}$, we can rerank the passages and obtain the most important passage within the document $d$ according to the similarity scores.

Our backbone models for the counterfactual learning method include two kinds of document retrieval models: (1) sparse document retrieval models: BM25~\cite{robertson2009probabilistic}, docT5query~\cite{nogueira2019doc2query}, deepCT~\cite{dai2020deepCT}; (2) dense document retrieval models: DPR~\cite{karpukhin2020dpr}, ANCE (FirstP)~\cite{xiong2020ance}, colBERT~\cite{DBLP:conf/sigir/KhattabZ20}, ME-BERT~\cite{luan2021mebert}.

\paragraph{\textbf{Adversarial Attack}}
We take four different attack methods for testing the robustness of the models: 
Term spamming (TS)~\cite{gyongyi2005web} randomly replaces words with terms randomly sampled from the target query.
(2) PRADA~\cite{wu2023prada} is a decision-based black-box ranking attack method against NRMs via word substitution.
(3) PAT~\cite{liu2022order} is an anchor-based ranking attack method against NRMs via trigger generation.
(4) MCARA~\cite{liu2023black} formalizes attacks on DR models as a contrastive learning problem in a multi-view representation space.

The baselines of the adversarial attack models are:
(1) Adversarial training (AT): We follow the vanilla AT method~\cite{goodfellow2014explaining} to directly include the adversarial examples during training.
(2) CertDR~\cite{wu2022certified} is a certified defense method for NRMs.
(3) PIAT~\cite{liu2023perturbation} is a novel perturbation-invariant adversarial training method.

\paragraph{\textbf{Evaluation Metrics}}
We evaluate the model's document retrieval performance by calculating the Mean Reciprocal Rank at the top 10 positions (MRR@10d) for the retrieved documents. 
Simultaneously, the model's ability to extract the key passages is assessed by calculating the Mean Reciprocal Rank at the top 10 positions, which is denoted using MRR@10p in this work, for the retrieved passages within the document. 
In particular, this metric quantifies the model's ability to rank relevant passages highly within the top 10 results. 
Statistical significance is tested using the permutation test with $p < 0.05$.

\subsection{Implementation}
We build our models with Pytorch~\cite{paszke2019pytorch} based on huggingface transformers~\cite{wolf2019huggingface}. 
Our systems use token dimension nt = 32 and  CLS dimension nc = 768 as default. 
All models are trained for 5 epochs with AdamW optimizer, a learning rate of 3e-6, 0.1 warm-up ratio, and linear learning rate decay, which takes around 12 hours. 
Hard negatives are sampled from top 1000 BM25 results.
Each query uses 1 positive and 7 hard negatives; each batch uses 12 queries on MSMARCOdocument. 
We conduct validation on randomly selected 512 queries from corresponding train set. Latency numbers are measured on dual Xeon E5-2630 v3 for CPU and RTX 3080 ti for GPU. 

The $\epsilon$ chosen for creating adversarial examples is 5\%.

\subsection{Research Questions}\label{ssec:RQs}
We aim to explore the following two research questions:
\begin{enumerate}[leftmargin=*]
    \item How can counterfactual learning methods contribute to the prediction of crucial segments within documents?
    \item How can counterfactual learning be incorporated into pretraining tasks to augment the capabilities of document retrieval models?
\end{enumerate}

These questions form the core of our investigation, delving into the potential impact of counterfactual learning on the identification of significant document segments, and its broader integration into the pretraining process to improve document retrieval model capabilities.

\begin{table*}[tb]
\renewcommand{\arraystretch}{1}
\caption{The retrieval effectiveness of retrieving passage within the document according to $\delta_{\text{rel}}$ with difference window size and different type of counterfactual construction methods. Window size of 128 reaches the best performance. The evaluation metric is MRR@10p. The best performance among various counterfactual document construction methods for a model is boldfaced. $\ast$ indicates significant improvements(p < 0.05).}
\label{5_couterfactual_type}
\begin{tabular}{cl ccc ccc }
\toprule
 & \multirow{2}{*}{ {Models}} 
 & \multicolumn{3}{c}{window size=128} & \multicolumn{3}{c}{window size=64} \\
 \cmidrule(lr){3-5}\cmidrule(lr){6-8}
&& \multicolumn{1}{p{1.8cm}<{\centering}}{deletion}       &  \multicolumn{1}{p{1.8cm}<{\centering}}{modification} & \multicolumn{1}{p{1.8cm}<{\centering}}{replacement} & \multicolumn{1}{p{1.8cm}<{\centering}}{deletion} &  \multicolumn{1}{p{1.8cm}<{\centering}}{modification} & \multicolumn{1}{p{1.8cm}<{\centering}}{replacement} \\ 
\midrule
 {\multirow{3}{*}{sparse}} & BM25       &  {\textbf{0.527}}* &  {0.501}        & 0.513       &  {0.491}    &  {0.476}        & 0.485       \\ 
 {}                        & docT5query &  {\textbf{0.568}}* &  {0.543}        & 0.552       &  {0.524}    &  {0.505}        & 0.517       \\ 
 {}                        & deepCT     &  {\textbf{0.575}}* &  {0.550}        & 0.559       &  {0.538}    &  {0.511}        & 0.523       \\ \midrule
 {\multirow{4}{*}{dense}}  & DPR        &  {\textbf{0.613}}* &  {0.597}        & 0.604       &  {0.589}    &  {0.565}        & 0.576       \\ 
 {}                        & ANCE       &  {\textbf{0.632}}* &  {0.609}        & 0.620       &  {0.606}    &  {0.587}        & 0.594       \\
 {}                        & colBERT    &  {\textbf{0.625}}* &  {0.604}        & 0.611       &  {0.601}    &  {0.581}        & 0.589       \\ 
 {}                        & ME-BERT    &  {\textbf{0.614}}* &  {0.595}        & 0.601       &  {0.591}    &  {0.568}        & 0.577       \\ 
\bottomrule
\end{tabular}
\end{table*}

\begin{table*}[tb]
\renewcommand{\arraystretch}{1}
\centering
\caption{The performance of retrieving passage within the document with different conterfactual learning methods. The shapley-value-based method demonstrates superior performance. The evaluation metric is MRR@10p. $\ast$ and $\dag$ indicates significant improvements over $\delta_{\text{rank}}$ and $\delta_{\text{rel}}$ (p < 0.05).}
\label{5_couterfactual_learning}
\begin{tabular}{cl p{2.1cm}<{\centering} p{2.1cm}<{\centering} p{2.1cm}<{\centering}p{2.1cm}<{\centering} p{2.1cm}<{\centering} p{2.1cm}<{\centering}}
\toprule
\multicolumn{2}{c}{\multirow{2}{*}{methods}}  & \multirow{2}{*}{$\delta_{\text{rank}}$} & \multirow{2}{*}{$\delta_{\text{rel}}$} & \multicolumn{2}{c}{shapley-value}      & \multirow{2}{*}{single passage} & \multirow{2}{*}{guideline} \\ \cmidrule(lr){5-6}
 &  &  &  & no-overlap & merge &   &  \\ 
\midrule
\multirow{3}{*}{sparse} & BM25  & 0.411  & 0.527 & 0.532  & \textbf{0.537}*$\dag$ & 0.374 & / \\ 
 & docT5query & 0.442  & 0.568  & 0.555  & \textbf{0.561}*$\dag$ & 0.501  & /  \\ 
 & deepCT     & 0.448  & 0.575  & 0.558  & \textbf{0.567}*$\dag$ & 0.497  & /   \\ 
 \midrule
\multirow{4}{*}{dense} & DPR    & 0.477  & 0.613 & 0.621 & \textbf{0.625}*$\dag$ & 0.605 & \underline{0.703} \\ 
& ANCE       & 0.486   & 0.632  & 0.633  & \textbf{0.638}*$\dag$ & 0.611 & \underline{0.725} \\ 
& colBERT    & 0.482   & 0.625  & 0.642  & \textbf{0.647}*$\dag$ & 0.623 & \underline{0.733}  \\ 
& ME-BERT    & 0.478   & 0.614  & 0.624  & \textbf{0.628}*$\dag$ & 0.608  & \underline{0.706}   \\ 
\bottomrule
\end{tabular}
\end{table*}

\section{Experimental Results}\label{sec:result}
In this section, we delve into the experimental results, findings, and the performance evaluation of both the counterfactual method and the contrastive task discussed in the earlier sections of the paper.

\subsection{Counterfactual Learning Methods}
To address \textbf{RQ1} - \textit{How can counterfactual learning methods contribute to the prediction of crucial segments within documents?},
we conduct experiments with various settings of counterfactual documents and different counterfactual learning methods.

\paragraph{\textbf{Counterfactual Document Construction}}
We explore the optimal settings for constructing the counterfactual documents. Table \ref{5_couterfactual_type} illustrates the results obtained with three types of counterfactual document construction methods and each is experimented with two different segmentation window sizes.
The $\delta_{\text{rel}}$ method results are showcased, with $\delta_{\text{rank}}$ and shapley exhibiting similar outcomes.

From Table\ref{5_couterfactual_type}, we observe that the performance of the models with
the window size of 64 performs poorly due to only 51.8\% of the positive passages remaining intact, leading to significant information loss and a reduction in the counterfactual document's quality.
In contrast, a window size of 128 preserves 95.8\% of positive passages, providing a conducive environment for the model to extract significant passages effectively. 
Among the three construction types, the deletion method consistently yields the best performance because deletion induces the most substantial information change. 
Conversely, the modification method, involving minor word substitutions, results in counterfactual documents too similar to the originals, thus showing the poorest performance. 

Consequently, we set the window size as 128 and the construction type as deletion for subsequent experiments.

\paragraph{\textbf{Key Passage Extraction Methods}}
We compare the performance of different key passage extraction methods. 
Table \ref{5_couterfactual_learning} displays the results concerning the retrieval of positive passages within documents. 
The ``single passage'' column involves treating the target passage as a short document and re-ranking passages directly based on relevance scores from document retrieval models. 
The ``guideline'' column represents testing the passage retrieval models trained with the MSMARCO-passage dataset, serving as a performance upper bound since these models are trained with the ground truth passage relevance labels.

The $\delta_{\text{rank}}$ method exhibits suboptimal performance, attributed to coarse rank differences that sometimes result in identical rank changes for different passages. 
This inadequacy provides an insufficient signal to discriminate passage importance. 
Furthermore, rank differences are susceptible to influence from other documents in the list, introducing additional noise to the score.
In contrast, the $\delta_{\text{rel}}$ method outperforms $\delta_{\text{rank}}$ by capturing relevance changes in the relevance score. 
Importantly, it is not influenced by other documents in the set, focusing solely on the current document.

The shapley-value-based method demonstrates superior performance, suggesting that it provide a better estimation of passage importance and capture more subtle differences between the original document and counterfactual document compared to $\delta_{\text{rel}}$.
This also indicates that our method of reducing computational complexity effectively estimates the Shapley value. 
Notably, the shapley-value-based method surpasses the single passage approach, highlighting the value of our counterfactual learning approach.

Between the two methods of Shapley value computation, the merge method performs better than the no-overlap method. 
This is because the no-overlap method only takes into account half of the documents, resulting in information loss.

\subsection{Counterfactual Contrastive Learning}
To address \textbf{RQ2} - \textit{How can counterfactual learning be incorporated into pretraining tasks to augment the capabilities of document retrieval models?}, we conduct experiments employing different strategies for loss weight and assess the model's robustness under various adversarial attack methods.

\begin{table*}[!htb]
\renewcommand{\arraystretch}{1}
\centering
\caption{The result of the key passage extraction of the counterfactual contrastive learning task. The coupling learning method yields the best performance. The evaluation metric is MRR@10p. $\ast$ indicates significant improvements over the original model (p < 0.05).}
\label{5_contrastive_learning}
\begin{tabular}{cl ccc ccc}
\toprule
& \multirow{2}{*}{loss weight strategy} & \multicolumn{3}{c}{MS-MARCO passage label}      & \multicolumn{3}{c}{gpt3.5   passage label}  \\ 
\cmidrule(lr){3-8} 
&   & \multicolumn{1}{p{1.8cm}<{\centering}}{$\delta_{\text{rank}}$} & \multicolumn{1}{p{1.8cm}<{\centering}}{$\delta_{\text{rel}}$} & \multicolumn{1}{p{1.8cm}<{\centering}}{shapley (merge)} & \multicolumn{1}{p{1.8cm}<{\centering}}{$\delta_{\text{rank}}$} & \multicolumn{1}{p{1.8cm}<{\centering}}{$\delta_{\text{rel}}$} & \multicolumn{1}{p{1.8cm}<{\centering}}{shapley (merge)} \\ 
\midrule
\multirow{3}{*}{$\text{DPR}_{counter}$}  & relevance score  & 0.486 & 0.619 & 0.629 & 0.474 & 0.613 & 0.623 \\ 
& shapley value  & 0.489 & 0.624 & 0.633 & 0.479 & 0.616 & 0.625 \\ 
& coupling learning  & \textbf{0.491}* & \textbf{0.628}* & \textbf{0.636}* & \textbf{0.483} & \textbf{0.619} & \textbf{0.629} \\ 
\midrule
\multicolumn{2}{c}{original DPR} & 0.477 & 0.613 & 0.625 & 0.477 & 0.613 & 0.625 \\
\midrule
\multirow{3}{*}{$\text{ANCE}_{counter}$} & relevance score & 0.489 & 0.640 & 0.641 & 0.483 & 0.630 & 0.635 \\ 
& shapley value & 0.493& 0.642 & 0.644 & 0.485 & 0.632 & 0.636 \\ 
& coupling learning & \textbf{0.499}* & \textbf{0.645}* & \textbf{0.649}* & \textbf{0.489} & \textbf{0.637} & \textbf{0.640} \\ 
\midrule
\multicolumn{2}{c}{original ANCE} & 0.486 & 0.632 & 0.638 & 0.486 & 0.632 & 0.638 \\
\bottomrule
\end{tabular}
\end{table*}

\begin{table}[]

\renewcommand{\arraystretch}{1}
\centering
\caption{The effect of the counterfactual contrastive learning to document retrieval, evaluated by MRR@10d and NDCG@10d. Our regularization improves key passage extraction without hurting document retrieval performance.}
\label{5_doc_retrieval}
\begin{tabular}{l c c}
\toprule
 & \begin{tabular}[c]{@{}c@{}}MSMARCO DOC \\ MRR@10\end{tabular} & \begin{tabular}[c]{@{}c@{}}TREC DL2019 DOC\\ NDCG@10\end{tabular} \\ \midrule
DPR & 0.277 & 0.554  \\ 
$\text{DPR}_{counter}$  & \textbf{0.280}   & \textbf{0.555} \\ \midrule
ANCE & 0.315   & 0.622  \\ 
$\text{ANCE}_{counter}$ & \textbf{0.317}  & \textbf{0.624}   \\ \bottomrule
\end{tabular}
\end{table}

\begin{table}[]
\renewcommand{\arraystretch}{1}
\centering
\caption{The ablation study of the performance on extracting the key passage. The evaluation metric is MRR@10p.}
\label{5_contrastive_learning_ablation}
\begin{tabular}{lccc}
\toprule
\multicolumn{1}{p{1.8cm}<{\centering}}{$L_{nat}$}                         & \multicolumn{1}{p{1.8cm}<{\centering}}{0.477} & \multicolumn{1}{p{1.8cm}<{\centering}}{0.613} & \multicolumn{1}{p{1.8cm}<{\centering}}{0.625}           \\ \midrule
$+L_{pos}$        & 0.466 & 0.598 & 0.609           \\
$+L_{neg}$        & 0.471 & 0.608 & 0.618           \\ 
$+L_{pos}+L_{neg}$ & 0.487 & 0.625 & 0.635           \\
$+L_{adv}$        & 0.478 & 0.613 & 0.627           \\ \midrule
All Losses                           & 0.491 & 0.628 & 0.636           \\ \bottomrule
\end{tabular}
\end{table}

\paragraph{\textbf{Retrieval Performance}}
Table \ref{5_contrastive_learning} depicts the performance of retrained models in key passage extraction, considering distinct loss weight strategies and various passage label settings. 
Rows labeled $\text{DPR}_{counter}$ and $\text{ANCE}_{counter}$ present the original MRR@10 of these two models (as shown in Table \ref{5_couterfactual_learning}).

Results from models trained with MSMARCO passage labels indicate that the coupling learning method ~\cite{liu2023perturbation} yields the best performance. This is attributed to its dynamic adjustment of hyperparameters during the training process. 
The observed improvement in extracting key passages suggests that our contrastive learning methods enhance the counterfactual learning ability of dense retrieval models.

Utilizing Shapley value as the loss weight outperforms using relevance scores as the loss weight, reinforcing that Shapley value better captures the local and global differences between $(d, d^*)$ compared to the relevance score.

Recognizing the challenge of obtaining high-quality passage-level relevance labels and the associated time and cost implications, we also conduct experiments generated by gpt-3.5~\cite{achiam2023gpt}.
In this scenario, relevant passages in the document are labeled by gpt-3.5 using a few-shot prompt. 
Results demonstrate that even with this training data, our method achieves a modest performance improvement, underscoring the effectiveness of our pretraining task.

Table \ref{5_doc_retrieval} presents the document retrieval performance of the retrained models. The results show slight improvement in document retrieval compared to the original models, though not statistically significant. 
This indicates that our contrastive learning task does not compromise the document retrieval ability while enhancing the capacity to extract key passages.

Table \ref{5_contrastive_learning_ablation} shows the ablation study of the performance on extracting the key passage. 
Results from only $+L_{pos}$ or only $+L_{neg}$ exhibit poor performance, as they only consider a part of the contrastive direction of the counterfactual document, significantly compromising the models' ability. 
The combination of $L_{pos}+L_{neg}$ yields the most significant improvement in performance, emphasizing that the cascaded comparison among $(d, d^*, d^-)$ can significantly enhance the models' ability to extract the key passage.
The result of $+L_{adv}$ shows no change in performance, indicating that this loss does not negatively impact the overall performance. Considering all the loss components together achieves the best performance in extracting key passages.

\begin{table*}[tb]
\renewcommand{\arraystretch}{1}
\centering
\caption{The retrained models' performance on extracting the key passage under the dense retrieval attack methods. Under all four attack methods, both DPR and ANCE show a lesser performance decrease after the contrastive learning task. The evaluation metric is MRR@10p.}
\label{5_adversarial_attack}
\begin{tabular}{lccccc}
\toprule& \multirow{2}{*}{Original documents} & \multicolumn{4}{c}{Attack   method} \\ \cmidrule(lr){3-6} 
& & \multicolumn{1}{p{2.7cm}<{\centering}}{TS} & \multicolumn{1}{p{2.7cm}<{\centering}}{PRADA} & \multicolumn{1}{p{2.7cm}<{\centering}}{PAT}  & \multicolumn{1}{p{2.7cm}<{\centering}}{MCARA}   \\
\midrule
DPR   & 0.613  & 0.584 (-4.7\%) & 0.562 (-8.3\%) & 0.565 (-7.8\%) & 0.549 (-10.4\%) \\ 
$\text{DPR}_{counter}$  & 0.626  & 0.603 (\textbf{-3.7\%}) & 0.587 (\textbf{-6.2\%}) & 0.588 (\textbf{-6.1\%}) & 0.576 (\textbf{-8.0\%})  \\ \midrule
ANCE  & 0.632  & 0.603 (-4.6\%) & 0.586 (-7.3\%) & 0.584 (-7.6\%) & 0.570 (-9.8\%)  \\ 
$\text{ANCE}_{counter}$ & 0.641 & 0.614 (\textbf{-4.2\%}) & 0.606 (\textbf{-5.5\%}) & 0.603 (\textbf{-5.9\%}) & 0.589 (\textbf{-8.1\%})  \\ 
\bottomrule
\end{tabular}
\end{table*}

\paragraph{\textbf{Model Robustness}}
Table \ref{5_adversarial_attack} presents the results of the retrained models' performance in extracting key passages under various adversarial attack methods. When compared with the original documents, those modified by the attack methods exhibit a decrease in model performance. Under all four attack methods, both DPR and ANCE show a lesser performance decrease after the contrastive learning task. Even under the state-of-the-art attacking method on dense retrieval models—MCARA, the retrained model demonstrates robustness. This suggests that our contrastive learning task improves the models' robustness in extracting key passages.

\begin{table}[]
\renewcommand{\arraystretch}{1}
\caption{The decrease of the retrieval performance under four adversarial attack methods. $\ast$ indicates significant improvements over other models (p < 0.05).}
\label{5_adversarial_attack_retrieval}
\resizebox{75mm}{!}{
\begin{tabular}{lcccc}
\toprule
\multicolumn{1}{p{1.cm}<{\centering}}{\multirow{2}{*}{model}} & \multicolumn{4}{c}{Percentage   decrease of the MRR@10d} \\ \cmidrule(lr){2-5} 
\multicolumn{1}{c}{}                       & TS           & PRADA        & PAT          & MCARA       \\
\midrule
AT                                         & -13.5\%     & -15.8\%     & -16.9\%     & -19.2\%    \\
CertDR                                     & -12.4\%     & -14.1\%     & -14.7\%     & -17.3\%    \\
PIAT                                       & -10.7\%     & -12.8\%     & \textbf{-12.5\%}*     & \textbf{-14.8\%}*    \\ \midrule
$\text{DPR}_{counter}$                            & \textbf{-10.1\%}     & \textbf{-12.3\%}     & -13.7\%     & -16.1\%  \\ 
\bottomrule
\end{tabular}}
\end{table}

\begin{table}[]
\renewcommand{\arraystretch}{1}
\centering
\caption{The ablation study of the robustness on extracting the key passage under the dense retrieval attack methods. }
\label{5_adversarial_attack_ablation}
\resizebox{75mm}{!}{\begin{tabular}{lcccc}
\toprule
\multirow{2}{*}{Training Loss} & \multicolumn{4}{c}{Percentage   decrease of the MRR@10p} \\ 
\cmidrule(lr){2-5} 
& TS & PRADA & PAT & MCARA \\ \midrule
$L_{nat}$   & -4.70\% & -8.30\% & -7.80\% & -10.40\% \\ 
$+L_{pos}$  & -7.10\% & -9.90\% & -9.30\% & -12.10\% \\ 
$+L_{neg}$  & -5.60\% & -9.00\% & -8.50\% & -11.10\% \\ 
$+L_{pos}+L_{neg}$  & -4.50\% & -7.70\% & -7.40\% & -9.80\%  \\ 
$+L_{adv}$  & -4.20\% & -7.10\% & -6.90\% & -9.10\%  \\
Total       & \textbf{-3.70\%} & \textbf{-6.20\%} & \textbf{-6.10\%} & \textbf{-8.00\%}  \\ 
\bottomrule
\end{tabular}}
\end{table}

Table \ref{5_adversarial_attack_retrieval} presents the robustness analysis, showcasing the decline in retrieval performance under four distinct adversarial attack methods. 
Notably, our counterfactual learning model consistently outperforms AT and CertDR across all attack methods, and it surpasses PIAT under TS and PRADA. 
This observation underscores the effectiveness of our counterfactual contrastive learning model in enhancing overall model robustness. 
PIAT performs exceptionally well under PAT and MCARA, tailored for intricate attack methods and requiring specific attack-generated training data. 
In contrast, our model attains comparable performance without the need for complex adversarial examples in the training dataset.

In Table \ref{5_adversarial_attack_ablation}, we conduct an ablation study on the robustness of key passage extraction under dense retrieval attack methods. 
The initial row represents the attack outcomes for the original DPR model, while subsequent rows depict results considering different components of the training loss.
Results solely from $+L_{pos}$ or $+L_{neg}$ exhibit subpar performance, focusing only on a segment of the contrastive direction within the counterfactual document, thereby compromising model robustness. 
The combination of $L_{pos}+L_{neg}$ demonstrates marginal robustness improvement, highlighting the positive contribution of the counterfactual contrastive loss. 
Importantly, the inclusion of $L_{adv}$ yields the most significant enhancement in robustness, underscoring the pivotal role of comparing the adversarial attack document with the counterfactual document. 
Comprehensive consideration of all loss components collectively achieves the optimum performance in extracting key passages under dense retrieval attack methods.
\section{Conclusion}
In conclusion, our research addresses critical challenges in dense retrieval models, focusing on key passage extraction and vulnerability to adversarial attacks. 
We have emphasized the importance of enhancing model sensitivity to fine-grained relevance matching signals as a key strategy to mitigate these challenges.

Our proposed counterfactual regularization methods aim to make dense retrieval models more interpretable and robust by increasing their sensitivity to modifications in key passages. 
We introduced a cooperative game theory-based counterfactual passage extraction method, which has demonstrated potential in enhancing model's robustness in real-world applications.

Experimentally, our approach shows promising results, revealing the model's capacity to learn key passages without explicit passage-level relevance annotations. 
More importantly, our regularized dense retrieval models exhibit superior resistance to adversarial attacks, outperforming existing anti-attack methods.

However, The efficiency and scalability of our methods need careful consideration when applied to large document collections. 
Handling the computational challenges for such large-scale deployments is an area that requires further research. 
Additionally, as adversarial attack techniques continue to evolve, our model may not be fully robust against new and sophisticated attacks. 
In the future, we plan to focus on these aspects to enhance the practicality and robustness of our models.

In light of these outcomes, our work provides a pathway to more interpretable and robust dense retrieval models. 
As we continue to grapple with the complexities of vast document corpora and evolving search demands, our research lays the groundwork for further improvements in model reliability and applicability.





%
\bibliographystyle{ACM-Reference-Format}
\balance
\bibliography{references}


\end{document}